# Effective Defect Prevention Approach in Software Process for Achieving Better Quality Levels

Suma. V., and T. R. Gopalakrishnan Nair

***Abstract*—**Defect prevention is the most vital but habitually neglected facet of software quality assurance in any project. If functional at all stages of software development, it can condense the time, overheads and wherewithal entailed to engineer a high quality product. The key challenge of an IT industry is to engineer a software product with minimum post deployment defects.

This effort is an analysis based on data obtained for five selected projects from leading software companies of varying software production competence. The main aim of this paper is to provide information on various methods and practices supporting defect detection and prevention leading to thriving software generation. The defect prevention technique unearths 99% of defects. Inspection is found to be an essential technique in generating ideal software generation in factories through enhanced methodologies of abetted and unaided inspection schedules. On an average 13 % to 15% of inspection and 25% - 30% of testing out of whole project effort time is required for 99% - 99.75% of defect elimination.

A comparison of the end results for the five selected projects between the companies is also brought about throwing light on the possibility of a particular company to position itself with an appropriate complementary ratio of inspection testing.

## I. Introduction

A defect in an application can lead to a harmful situation in all phases of software development process. Anything connected to defect is a continual process and not a state. Defect prevention activity stems from comprehension of defects. A defect refers to any inaccuracy or blemish in a software work product or software process. The term defect refers to an error, fault or failure [1]. The IEEE/Standard defines the following terms as Error: a human action that leads to incorrect result.

Fault: incorrect decision taken while understanding the given information, to solve problems or in implementation of process. A Failure: inability of a function to meet the expected requirements [2]–[3].

Suma V. is with Information Science and Engineering Department, Dayananda Sagar College of Engineering, Bangalore, India (e-mail:sumavdsce@gmail.com).

T. R. Gopalakrishnan Nair is with Research and Industry Incubation Center, Dayananda Sagar Institutions, Bangalore, India (e-mail: trgnair@ieee.org).

Defect prevention [DP] is a process of identifying defects, their root causes and corrective and preventive measures taken to prevent them from recurring in future, thus leading to the production of a quality software product [4]-[5]-[11]-[12]-[15]. Hence, organizations should opt for defect detection and prevention strategies for long-term Return on Investment (ROI).

Among several approaches, inspection has proven to be the most valuable and competent technique in defect detection and prevention [5]-[13]-[14]-[15]. Identified defects are classified at two different points in time 1) time when the defect was first detected and 2) time when defect got fixed. Orthogonal Defect Classification (ODC) is the most prevailing technique for identifying defects wherein defects are grouped into types rather than considered independently. This technique highlights those areas in software development process that require attention [6]–[14].

If a defect dwells for a longer time in the product, it is more expensive to fix it. Therefore, it is necessary to reduce defect injection and boost defect removal efficiency. Defect removal efficiency (DRE) metric quantifies the excellence of the product by computing the number of defects before release of the product to the total number of latent defects [7]-[17].

DRE = number of defects removed during development phase / total number of latent defects

DRE depends upon time and method used to remove defects. But it is always more lucrative for defects to be prevented rather than detected and eliminated.

Certain amount of defects can be prevented through error removal techniques like educating development team through training, by use of formal specifications and formal verifications. It can also be prevented with use of tools, technologies, process and standards. Several tools are available right from requirements phase to maintenance phase to automate the entire development process. Usage of object oriented technology reduces interaction problems thus reducing number of defects arising in these areas. Defects can be prevented with the choice of appropriate process and in compliance with the process. By inculcating quality standards in software development, defects can be prevented to a maximum extent. Root cause analysis for defects is identified to be very successful in prevention of defects in all the booming software companies [8]-[9]-[16].

## II. CASE STUDY

The following case studies provide information on various defect detection and prevention techniques that are incorporated in mature companies in delivering a high quality product. This also includes one company that does not strictly adhere to DP strategies.

### *A. Effective Defect Prevention Techniques Adopted in a Leading Product Based Company in Embedded Systems*

The company follows staged process model, which is a representation of CMMI Meta model. The staged process defines five maturity levels and the process areas that are required to achieve a maturity level.

Since 1999-2000, the company follows qualitative and quantitative analysis as a defect preventive strategy. A data base is maintained to capture the mistakes identified after product is shipped to the field. Qualitative analysis comprises of stage kick off meeting to be carried out prior to the start of each life cycle phase or task to highlight those areas where mistakes were committed, identified and actions that were taken for their rectification in the past. Sensitization and discussions are carried out for the current project by handing over the documents of the lessons learned from previous similar type of projects. The core intention is to reduce defect injection and increase defect removal efficiency [10].

In quantitative approach, authentic and realistic data are collected from the stored projects. Based on 80% rule, projects are categorized on platform and technology upon which they were implemented. Control charts are used to inspect for consistency checks at all phases of software development life cycle. If an inspection at a phase exemplifies the non-conformance of the defects in the control band, it reveals the fact that either review was excellent or if review was reprehensible.

Testing comprises of

- Regression testing which ensures non introduction of unintentional behavior or additional errors in the software
- Performance test is conducted to ascertain the performance of requirements.
- Environmental test ensures testing of environment in which the product is to be deployed.
- Health test is also conducted for users of the product in compliance with health safety standards.

The review efficiency metric gives an insight on quality of review conducted. Review efficiency is idyllic if it can identify *one critical defect* per *every one man* hour spent on reviews.

Review Efficiency = Total number of defects found by reviews /Total number of defects in product

With a review efficiency of 87%, the company reported increasing their productivity from 250 to 400 accentuating the importance of adopting DP strategies. With an inspection-testing time ratio of 15:30, the company was able to reach a quality level of 99.75% defect-free product.

***Observation***

Inspection is carried out at all phases of software development rather than performing it only at coding stage. Inspection is carried out for requirement specifications, high level design and low level design in addition to code reviews. Company schedules 15% of the total time of the project for inspections and 30% for testing to achieve a quality of 99.75% defect-free product.

As the deployable product is almost free from defects, cost entailed for rework is quite nominal. Since cost of fixing defects after shipment is 10 times more than the cost of fixing it in-house [9], inspection becomes mandatory for highly safety critical systems [11].

Table 1 depicts the estimated time and actual time for 5 different projects. Average time estimated for inspection is 13.9 % of total project time and actual time taken is 14.2%. Test time is estimated to be 28.2% but actual test time is 30.8% of the total project time. Thus for highly critical systems an inspection of 15% and testing of 30% is good enough to achieve nearly 99.75 % of defect-free product.

### *B. Effective Defect Prevention Techniques Adopted in a Leading Service Based Software Company*

The company follows staged continuous model, which is a representation of CMMI Meta model. The continuous process defines five capability levels and process areas that are assessed for five capability levels.

Since 2002, the company is adhering to defect detection and defect prevention techniques to enhance quality of the product. The defect detection techniques are review of plans, schedules, and records. Product and process audits are carried out as part of quality control to uncover defects and correct them. The defect prevention techniques followed in the company includes pro-active, reactive and retrospective DP.

Pro-active DP is to create an environment for controlling defects rather than just reacting to it. A stage kick off meeting is conducted to reveal those areas where slip-ups were committed, recognized and actions taken for their refinement in the past. Company considers from the previous projects, the lessons learnt from the life cycle phases, the DP action items documented and best practices adopted. It leverages from other projects, the DP action items in the organization that are of same nature.

Reactive DP identifies and conducts RCA (Root Cause Analysis) for defects meeting at trigger points or at logical points. Therefore, curative actions are employed along with preventive actions to eliminate the potential defects. The most common root causes for defects identified are lack of communication, lack of training, oversight, project methodology and planning.

Retrospection is performed towards the end of the project or at identified phases to identify strong points and to explore the areas requiring perfection.

***Observation***

Inspection is executed at all stages of software development. Testing activity includes automation testing, verification and validation testing. From 5 projects listed in and actual time taken is 14.7% of total project time. With an estimate of 25% of testing time, it is seen that 25.1% of total development time is required to achieve nearly 99% defect-free software product.

The cost of rework for 1% of defect when identified at the customer's site is 10 times the cost required for fixing the same defect when identified in-house [11]. As a matter-of-fact, companies adopting to DP strategies have shown that over a period of time, quality of the product is enhanced while the cost of quality is reduced [12].

also been studied. Because DP is not stringently followed, a substantial amount of time is spent on developer unit testing,

***Observation***

The company schedules 5% of total project time for inspection which necessitates almost 40% of testing time out of the total development effort. Defects that can be captured with this ratio of inspection and testing are only 80%.

Cost required for rework is found to be more expensive than the cost incurred in adhering to DP strategies. Of the selected 5 projects, it is observed from the Table III that with for inspection is 5.5%. This ratio requires a scrupulous testing of 40.6% of actual time against an estimated time of 34.5% out of whole project time. If inspection time is increased in all phases of software development life cycle, then testing time gets significantly reduced.

Fig. 1 shows a comparative graph of inspection and testing for 5 selected projects from 3 different companies. From the graph, it is clear that with increase in inspection time, testing time gets decreased, as most of the defects get uncovered during inspection. Investment in inspection is initially high but over a period of time it becomes stable, which means cost is reduced and quality is increased.

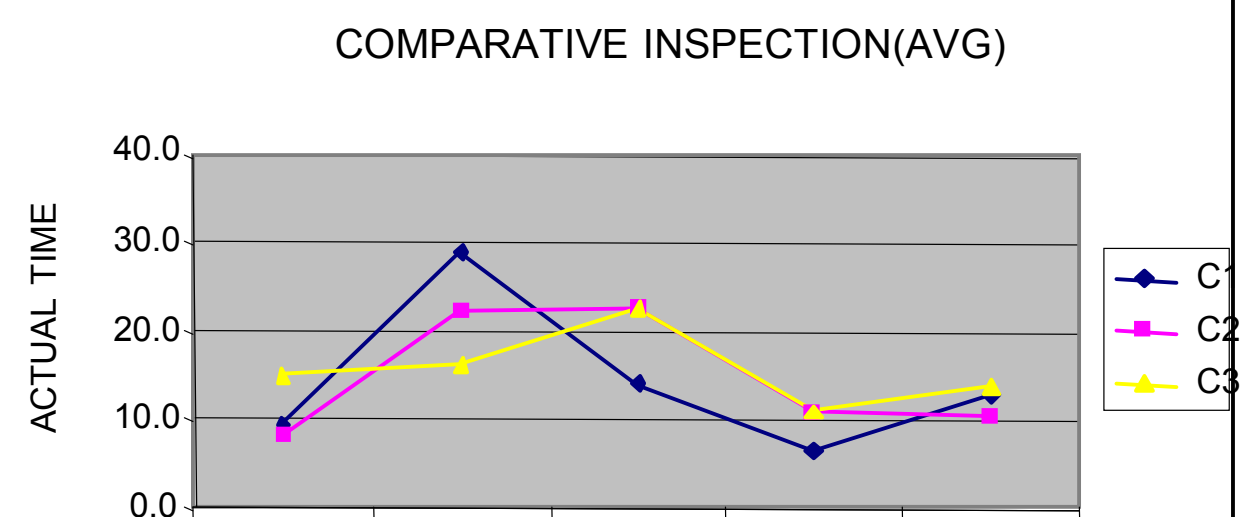

| | P1 | P2 | P3 | P4 | P5 |
|---|---|---|---|---|---|
| C1 | 9.3 | 29.0 | 13.7 | 6.3 | 12.7 |
| C2 | 8 | 22 | 22.3 | 10.7 | 10.3 |
| C3 | 15.0 | 16.0 | 22.3 | 10.7 | 13.7 |

PROJECTS

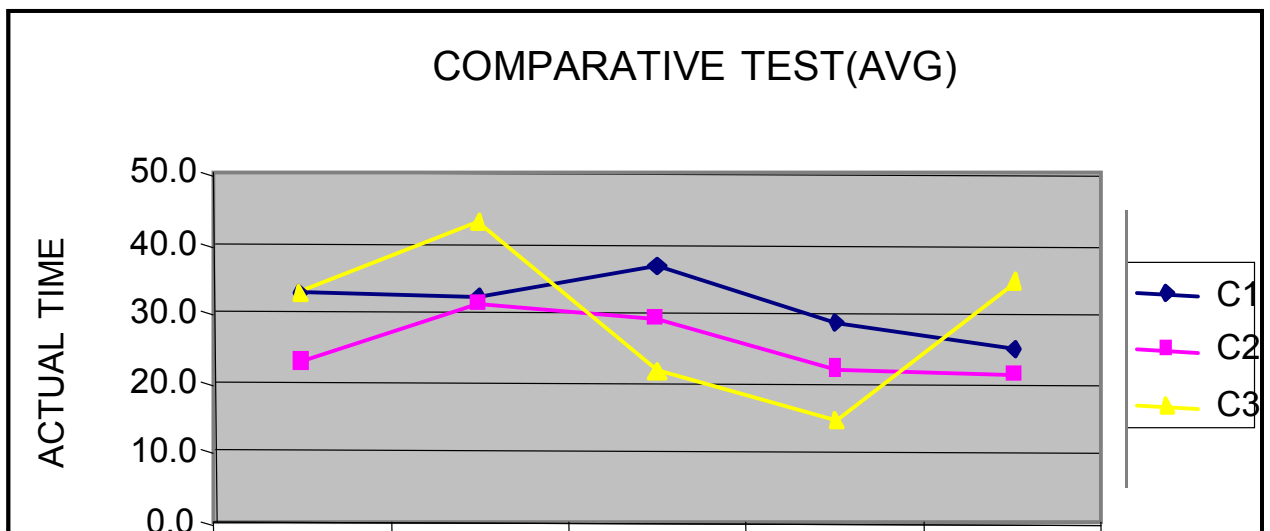

| | P1 | P2 | P3 | P4 | P5 |
|---|---|---|---|---|---|
| C1 | 32.7 | 32.0 | 36.3 | 28.3 | 24.7 |
| C2 | 22.7 | 31.3 | 29.0 | 21.7 | 21.0 |
| C3 | 32.7 | 42.8 | 21.3 | 14.0 | 33.9 |

PROJECTS

Fig. 1 Comparative graphs of inspection and testing for 3 companies over 5 selected projects

## III. CONCLUSION

Implementation of defect prevention strategy not only reflects a high level of process maturity but is also a most valuable investment. The detection of errors in development life cycle helps to prevent the passage of errors from requirement specification to design and from design to code. Analysis carried out across three companies shows the importance of incorporating defect prevention techniques in delivering a high quality product. The focal point of quality cost investment is to invest in right DP activities rather than investing in rework which is seen as an outcome of uncaptured defects. There are several methods, techniques and practices for defect prevention. Software inspection has proved to be the most successful defect detection and prevention technique. The goal of reaching a consistently 99 % defect-free software depends much on effective defect prevention techniques adopted.

## REFERENCES

[1] Brad Clark, Dave Zubrow, "How Good Is the Software: A review of Defect Prediction Techniques", sponsored by the U.S. department of Defense 2001 by Carnegie Mellon University, version 1.0, pg 5.

[2] The Software Defect Prevention /Isolation/Detection Model drawn from www.cs.umd.edu/~mvz/mswe609/book/chapter2.pdf

[3] Jeff Tian "Quality Assurance Alternatives and Techniques: A Defect-Based Survey and Analysis", ASQ by Department of Computer Science and Engineering, Southern Methodist University , SQP Vol. 3, No. 3/2001.

[4] Purushotham Narayan, “Software Defect Prevention in a Nut shell”, Copyright © 2000-2008 iSixSigma LLC. See also software.isixsigma.com/library/content/c030611a.asp - 73k –

[5] S.Vasudevan, “Defect Prevention Techniques and Practices” proceedings from 5th annual International Software Testing Conference in India, 2005.

[6] Chillarege, I.S. Bhandari, J.K. Chaar, M.J. Halliday, D.S. Moebus, B.K. Ray, M.-Y. Wong, "Orthogonal Defect Classification-A Concept for In-Process Measurements," IEEE Transactions on Software Engineering, vol. 18, no. 11, pp. 943-956, Nov., 1992 .

[7] Craig Borysowich , “Inspection/Review Meeting Metrics”, 2006. See also blogs.ittoolbox.com/eai/implementation/archives/sample-inspectionreview-metrics-13640 - 184k

[8] Halling M., Biffl S. (2002) "Investigating the Influence of Software Inspection process Parameters on Inspection Meeting Performance", Int. Conf. on Empirical Assessment of Software Engineering (EASE), Keele, April 2002.

[9] Stefen Biffl, Michael Halling, “ Investingating the Defect Detection Effectiveness and Cost Benefit of Nominal Inspection Teams “, IEEE Transactions On Software Engineering, Vol 29, No.5, May 2003

[10] Defect Prevention by SEI’s CMM Model Version 1.1.,, http://ww.dfs.mil/techlogy/pal/cmm/lvl/dp.

[11] Watts S. Humphrey, "Managing the Software Process", Chapter 17 – Defect Prevention, ISBN-0-201-18095-2

[12] Kashif Adeel, Shams Ahmad, Sohaib Akhtar, “Defect Prevention Techniques and its Usage in Requiremnts Gathering-Industry Practices”, Paper appears in Engineering Sciences and Technology, SCONEST, ISBN 978-0-7803-9442-1, pg 1-5,August 2005

[13] Joe Schofield, “Beyond Defect Removal: Latent Defect Estimation with Capture Recapture Method (CRM)”, published in IT Metrics and Productivity Journal, August 21, 2007

[14] Adam A. Porter, Carol A. Toman and Lawrence G Votta, “An Experiment to Assess the Cost-Benefits of Code Inspections in Large Scale Software Development”, IEEE Transactions on Software Engineering, VOL. 23, NO. 6, June 1997

[15] Lars M. Karg, Arne Beckhaus, “ Modelling Software Quality Costs by Adapting Established Methodologies of Mature Industries", Proceedings of 2007 IEEE International Conference in Industrial Engineering and Engineering Management in Singapore, ISBN 078-1-4244-1529-8, Pg 267-271, 2-4 Dec.2007

[16] David N. Card, “Myths and Stratergies of Defect Causal Analysis”, Proceedings from Pacific Northwest Software Quality Conference, October.

[17] K.S. Jasmine, R. Vasantha ,”DRE – A Quality Metric for Component based Software Products”, proceedings of World Academy Of Science, Engineering and Technonolgy, Vol 23, ISSN 1307-6884, August 2007.

TABLE I
TIME AND DEFECT PROFILE OF A LEADING SOFTWARE PRODUCT BASED COMPANY

| | P1 | | P2 | | P3 | | P4 | | P5 | |
|---|---|---|---|---|---|---|---|---|---|---|
| Total time ( in man hours) | 250 | 263 | 200 | 201 | 340 | 355 | 170 | 167 | 100 | 101 |
| Req time | 20 | 23 | 30 | 35 | 25 | 23 | 15 | 14 | 23 | 25 |
| Req review | 2 | 1 | 5 | 6 | 3 | 3 | 2 | 1 | 2 | 2 |
| Req test | 5 | 6 | 8 | 8 | 5 | 6 | 5 | 7 | 4 | 5 |
| Design time | 35 | 40 | 80 | 82 | 45 | 46 | 28 | 29 | 40 | 46 |
| Design review | 3 | 3 | 10 | 14 | 6 | 5 | 5 | 3 | 5 | 5 |
| Design test | 9 | 11 | 20 | 22 | 9 | 10 | 8 | 10 | 8 | 11 |
| Implementation time | 90 | 100 | 200 | 201 | 115 | 118 | 52 | 50 | 100 | 101 |
| Code review | 15 | 16 | 26 | 27 | 15 | 17 | 9 | 7 | 15 | 17 |
| Testing | 78 | 81 | 64 | 66 | 90 | 93 | 60 | 68 | 50 | 58 |
| Insp Avg | 10 | 9.3 | 23.7 | 29 | 14 | 13.7 | 10 | 6.3 | 12 | 12.7 |
| Test Avg | 30.7 | 32.7 | 30.7 | 32 | 34.7 | 36.3 | 24.3 | 28.3 | 20.7 | 24.7 |

Shaded column indicates estimated values and unshaded columns indicate the actual values
Req – Requirement, Insp – Inspection, Avg – Average

TABLE II
TIME AND DEFECT PROFILE OF A LEADING SERVICE BASED COMPANY

| | P1 | | P2 | | P3 | | P4 | | P5 | |
|---|---|---|---|---|---|---|---|---|---|---|
| Total time (in man hours) | 250 | 263 | 502 | 507 | 340 | 368 | 166 | 167 | 255 | 263 |
| Req time | 25 | 20 | 50 | 55 | 32 | 32 | 10 | 12 | 25 | 20 |
| Req review | 4 | 5 | 10 | 12 | 12 | 13 | 5 | 7 | 6 | 7 |
| Req test | 4 | 4 | 10 | 9 | 12 | 10 | 5 | 5 | 7 | 3 |
| Design time | 40 | 45 | 100 | 110 | 40 | 45 | 20 | 22 | 40 | 45 |
| Design review | 6 | 6 | 18 | 20 | 17 | 19 | 8 | 11 | 8 | 9 |
| Design test | 7 | 8 | 19 | 17 | 18 | 12 | 10 | 10 | 9 | 8 |
| Implementation time | 85 | 100 | 180 | 165 | 100 | 105 | 45 | 40 | 85 | 100 |
| Code review | 12 | 13 | 32 | 34 | 31 | 35 | 13 | 14 | 12 | 15 |
| Testing | 55 | 56 | 63 | 68 | 61 | 65 | 45 | 50 | 50 | 52 |
| Avg insp | 7.3 | 8 | 20 | 22 | 20 | 22.3 | 8.7 | 10.7 | 8.7 | 10.3 |
| Avg test | 22 | 23 | 30.7 | 31.3 | 30.3 | 29 | 20 | 21.7 | 22 | 21 |

TABLE III
TIME AND DEFECT PROFILE OF A COMPANY NOT STRINGENT TO DP

| | P1 | | P2 | | P3 | | P4 | | P5 | |
|---|---|---|---|---|---|---|---|---|---|---|
| Total time (in man hours) | 225 | 230 | 490 | 507 | 340 | 368 | 150 | 159 | 240 | 250 |
| Req time | 20 | 24 | 54 | 55 | 28 | 30 | 15 | 19 | 30 | 30 |
| Req review | 2 | 2 | 3 | 4 | 3 | 3 | 1 | 1 | 2 | 2 |
| Testing | 9 | 10 | 20 | 22 | 16 | 16 | 6 | 7 | 10 | 11 |
| Design time | 30 | 35 | 70 | 77 | 40 | 42 | 30 | 33 | 45 | 45 |
| Design review | 3 | 4 | 4 | 5 | 3 | 3 | 1 | 1 | 2 | 3 |
| Testing | 11 | 12 | 26 | 28 | 17 | 18 | 9 | 10 | 13 | 17 |
| Implementation time | 85 | 100 | 180 | 165 | 100 | 105 | 45 | 40 | 85 | 100 |
| Code review | 6 | 6 | 13 | 13 | 12 | 13 | 9 | 10 | 12 | 12 |
| Testing | 68 | 80 | 120 | 133 | 93 | 105 | 50 | 68 | 65 | 72 |
| Avg insp | 3.7 | 4 | 6.7 | 7.3 | 6 | 6.3 | 3.7 | 4 | 4 | 5.7 |
| Avg test | 29 | 34 | 55.3 | 61 | 42 | 46.3 | 21.7 | 28.3 | 29.3 | 33.3 |